\documentclass{iaus}
\usepackage{graphicx}

\title[Properties of Post-AGB Stars] 
{Properties of Post-AGB Stars}

\author[P. Garc\'\i a-Lario]   
{P. Garc\'\i a-Lario$^{1,2}$}

\affiliation{$^1$ISO Data Centre.
European Space Astronomy Centre. Research
and Scientific Support Department of ESA. Madrid, Spain
\break email: Pedro.Garcia.Lario@esa.int\\
$^2$Herschel Science Centre. 
European Space Astronomy Centre. Research
and Scientific Support Department of ESA. Madrid, Spain}

\pubyear{2006}
\volume{234}
\jname{Planetary Nebulae in our Galaxy and Beyond}
\editors{M. J. Barlow \& R. H. M\'endez, eds.}

\begin{document}

\maketitle

\begin{abstract}
  A review is presented of the most relevant results obtained in the last 
 few years on this rare class of astronomical sources.
   Multi-wavelength analysis of an increasing number of post-AGB stars
 reveal that they constitute a more inhomogeneous population of stars 
 than previously thought. The new data available allow us to study 
 these sources with unprecedent spatial resolution 
 and to extend our spectroscopic knowledge in a 
 systematic way  to the infrared for the first time, where crucial information
 is contained on 
 the chemical composition of the gas and dust in their  circumstellar shells.
  The overall infrared properties derived from ISO and Spitzer data can  
 be used to trace the mass loss history and the chemical evolution of the
 ejected material.  The new results impose severe observational
 constraints to the current nucleosynthesis models and suggest that the
 evolution is mainly determined not only  by the initial mass but also by
 the metallicity of the progenitor star.
   Post-AGB samples are likely to grow in the near future with the
 advent of new data from space facilities like Spitzer or Akari. 
 Studies of post-AGB stars in the galactic halo, the Magellanic Clouds and
 other galaxies of the Local Group will certainly improve our knowledge on the 
 evolutionary connections between AGB stars and PNe.
\end{abstract}

\firstsection 
\section{Introduction}

  Post-Asymptotic Giant Branch (post-AGB, hereafter) stars are 
low- and intermediate-mass (1--8 M$_{\odot}$) stars in the transition 
from the AGB to the planetary nebula (PN, hereafter) stage. 
This phase of stellar evolution has recently been discussed in detail
in excellent reviews by van Winckel (2003) and by Waelkens \& Waters (2004),
to which the interested reader is referred. In spite of the progress made in
the last few years, post-AGB stars, considered as a class, are still far 
from being completely understood. Two basic reasons exists for this: first,
the number of sources which at any given time are evolving along this
short-lived (10$^2$$-$10$^3$ yr) phase in our Galaxy is very small; 
second, in many cases their evolution takes place completely hidden from our
view, as these stars develop thick circumstellar envelopes
during the previous thermal pulsing AGB phase, which makes observations
in the optical very difficult, if not impossible.

 Stars enter the post-AGB 
at the moment when the strong AGB mass loss stops, which is also 
accompanied by the end of the stellar pulsations. At this moment, the central 
star is expected to evolve to higher effective temperatures, 
reappearing again in the optical range as a consequence of the dilution of 
the circumstellar envelope on a timescale which is mainly dependent on the
initial mass of the progenitor star (Bl\"ocker 1995).
 Althoug a basic knowledge of the whole process exists, there are  
still many open issues which deserve further investigation.  
The mechanism(s) of PN shaping, the dual-dust chemistry phenomenon observed
not only in transition sources but also later during the PN stage, or
the recent identification of some solid state features which are only
detected in post-AGB stars whose carriers are
still unknown, are just a few examples of areas where there is still
progress to be made, some of which will be here addressed.

\section{Recent post-AGB surveys}

\subsection{New identifications, atlases and catalogues}

 In the last decade a considerable effort has been made in the 
identification of the missing population of post-AGB stars
which may have escaped detection in the old searches carried out 
in the early 90's. Most of these surveys, although based upon the 
characteristic excess displayed by these transition sources at IRAS
wavelengths, were strongly biased toward candidates showing bright 
optical counterparts,
usually stars located at relatively high galactic latitudes and, as such, 
belonging to a low-mass population (e.g. Hrivnak et al. 1989; Oudmaijer et al. 
1992; Hu et al. 1993; Oudmaijer 1996).

  This is not the case of the GLMP catalogue (Garc\'\i a-Lario 1992),
a colour selected, flux limited sample of IRAS sources which contains more 
than 250 candidate post-AGB stars, the largest compilation so far available. 
Actually, only half of these sources show optically bright counterparts, for
which low resolution spectroscopy, finding charts and improved astrometric 
coordinates have recently been compiled in an atlas by Su\'arez et al. (2006).
The rest are heavily obscured stars only detectable in the infrared.

  The spectral type distribution shown by the newly classified
sources do not show the strong peak at intermediate F-G classes
typically observed in {\it classical} post-AGB samples. In contrast, they
display a much flatter distribution of spectral types. Remarkably, the 
subsample of extremely obscured sources showing no optical counterpart 
follows a galactic distribution which corresponds to a more massive
population (see Figure 1).

\begin{figure}[b!]
\centering
 \resizebox{6.3cm}{!}{\includegraphics{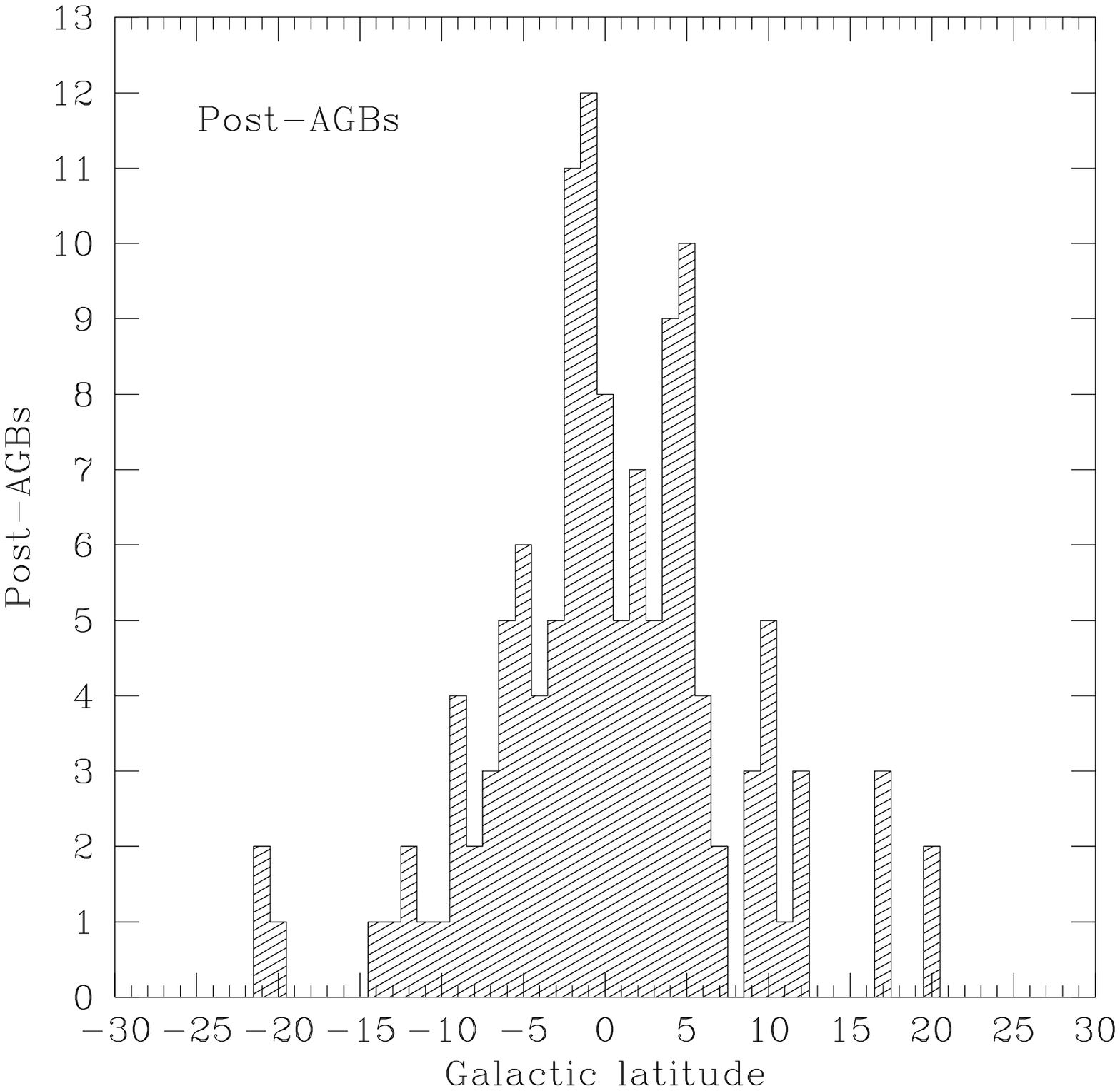} }
 \resizebox{6.3cm}{!}{\includegraphics{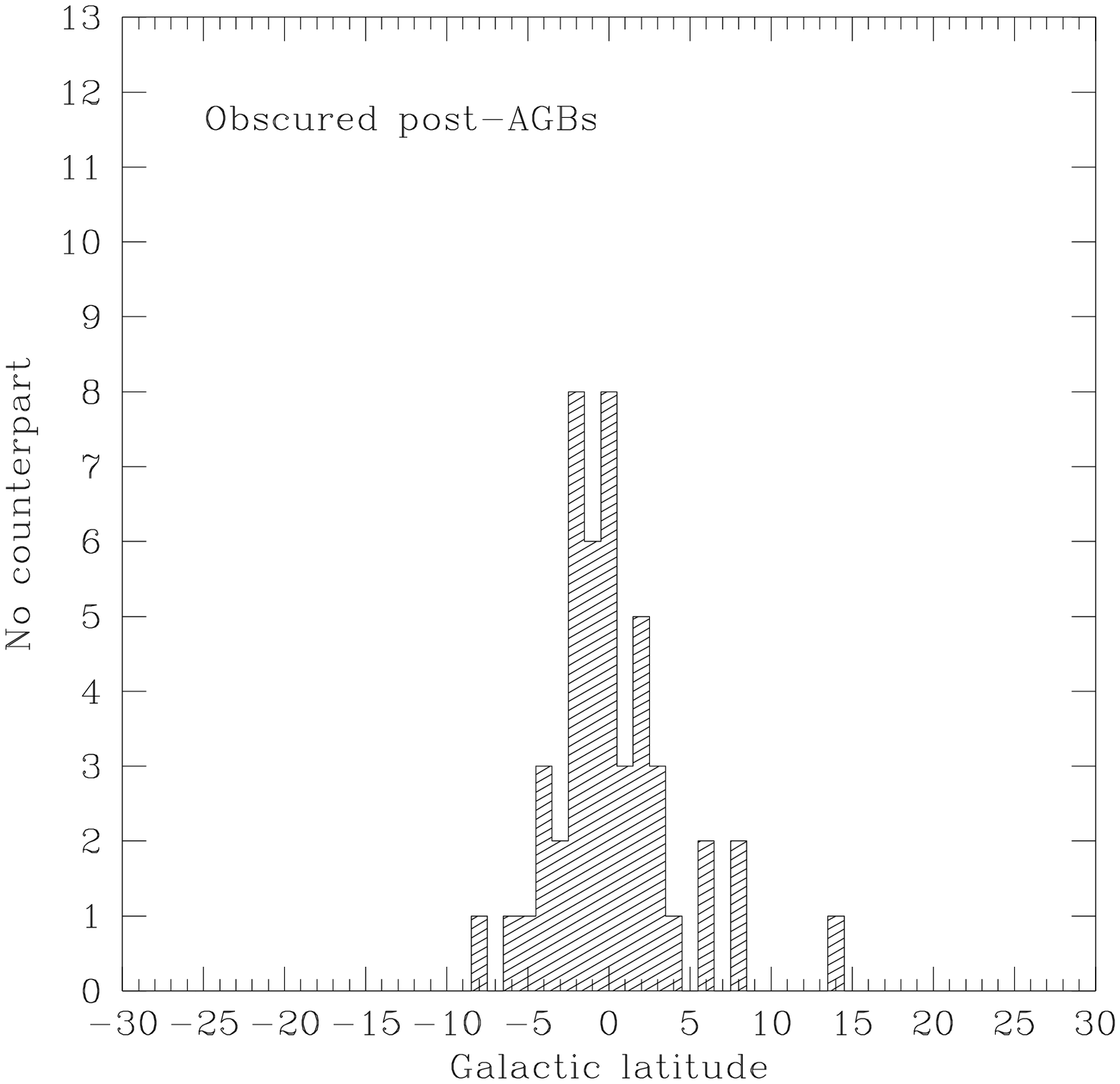} }
  \caption{Galactic latitude distribution of {\it classical}, optically
 bright post-AGB stars (left) and of the new population of obscured post-AGB 
 stars (right). Note the narrower distribution shown by the 
 obscured sample, which suggests a higher mass population (from  Su\'arez et 
 al. 2006).}\label{Orichsequence}
\end{figure}

 More recently, Bayo et al. (in prep.) have used the improved 
sensitivity and spatial resolution of the MSX galactic plane survey,
 combined with some of the more recent tools
developed for the Astrophysical Virtual Observatory project
({\tt http://www.euro-vo.org}) to carefully cross-match the MSX and IRAS 
Point Source Catalogues. This way some 100 additional
sources have been identified as new, low-galactic latitude post-AGB stars,   
for which nothing is yet known.

  In parallel, Szczerba et al. are building a catalogue 
where all the information available in the 
literature on stellar parameters and chemical abundances of post-AGB stars
will be collected. Special 
attention is also given to the infrared spectral features detected 
by ISO and to the morphology inferred from HST images 
(Si\'odmiak et al., these proceedings).

\subsection{Morphological surveys}

 Studying the early shaping of PNe through morphological surveys with 
subarcsec resolution is another area where important progress has been made 
in the last few years, although
no conclusion has yet been reached on the physical mechanism(s) which 
lead to the generation of the macro- and micro-structures later
observed during the PN phase (Corradi, these proceedings). Interaction 
with a binary companion (Livio \& Soker 1988) and magnetic fields 
(Garc\'\i a-Segura et al. 1999) continue to be the most popular options. 

 Following the pioneering survey made with HST/WFPC2 by Ueta et al. (2000),
other HST snapshot surveys have been carried out using 
ACS and NICMOS (see e.g. Sahai et al., these proceedings),
extending the observational database in size and wavelength coverage. 
The results obtained confirm that almost all post-AGB stars  
are aspherical, sometimes showing complex multipolar morphologies
in reflected light.  

Additional observations are 
now being collected from the ground, taking adavantage of the 
new instrumentation and innovative observing techniques made available at 
8 m class telescopes, which are able to compete with the spatial resolution 
obtained from space with HST, especially at near- and mid-infrared
wavelengths, where the detailed morphologies of the dusty waists are better
probed. Nice examples were shown at this conference, like the Keck 
images of a sample of proto-PNe obtained using laser guide star adaptive 
optics at near infrared wavelengths (S\'anchez Contreras et al.); or
the VLT interferometric measurements of the 
peculiar AGB star OH 231.8+4.2 made 
with MIDI at mid-infrared wavelengths and with the adaptive optics system 
NACO in the near-infrared (Matsuura et al.). 
Subarcsec resolution has also been achieved at mid-infrared wavelengths
from the ground using OSCIR at the infrared-optimized Gemini telescopes 
(Kwok et al. 2002; Clube \& Gledhill 2004, these proceedings).

\subsection{Molecular hydrogen surveys}

  Data collected in the last few years include subarcsecond resolution 
images of individual sources using integral field spectroscopy  (Lowe 
\& Gledhill,  these proceedings) as well as low- and high-resolution 
near-infrared spectroscopy of different samples of post-AGB stars 
(Garc\'\i a-Hern\'andez et al. 2002; Kelly \& Hrivnak 2005; Hrivnak, 
these proceedings). 
The observations suggest that the fluorescence-excited molecular hydrogen 
emission is the result of the central star becoming hot 
enough to produce the necessary UV photons while, in contrast, shock-excited
emission is detected at any moment during the post-AGB phase, irrespective of 
the effective temperature of the central star, 
in association with bipolar outflows, as a consequence of the 
interaction of the new, fast post-AGB wind with the older, slow AGB  wind. 

\section{Multi-wavelength analysis of individual sources}

\subsection{Properties of classical, optically bright post-AGB stars}

{\it Classical} post-AGB stars are usually considered to be slowly
evolving, low-mass 
stars belonging to the old disk population. This is mainly because
they show low metallicities (typically from [Fe/H] = $-$1 to [Fe/H] = $-$0.3),
and a high galactic latitude distribution.
The overall SED shows a characteristic double-peak component (see Figure 2), 
the optical peak corresponding to the bright central star
while in the infrared we can see the emission from the 
cold dust grains in the circumstellar envelope. 
 In the optical, they  are affected by little to moderate 
reddening and when observed with HST they usually show only slightly aspherical 
morphologies which are detected in scattered light, although in a few cases 
more complex multipolar morphologies are also observed. 

 They can be divided in two main subgroups: C-rich sources, which 
show strong s-process enrichment indicative of an efficient 3rd dredge-up
usually accompanied by a 21 $\mu$m emission;
and O-rich sources, which are not s-process enriched.
The latter group may be the result of the evolution of AGB stars with 
very low progenitor masses
which do not experience an efficient 3rd dredge-up (van Winckel 2003).

\begin{figure}[h!]
\centering
 \resizebox{6cm}{!}{\includegraphics[height=5cm,width=4.5cm]{garcialariof2left.ps} }
 \resizebox{6cm}{!}{\includegraphics[height=4.5cm,width=4.5cm]{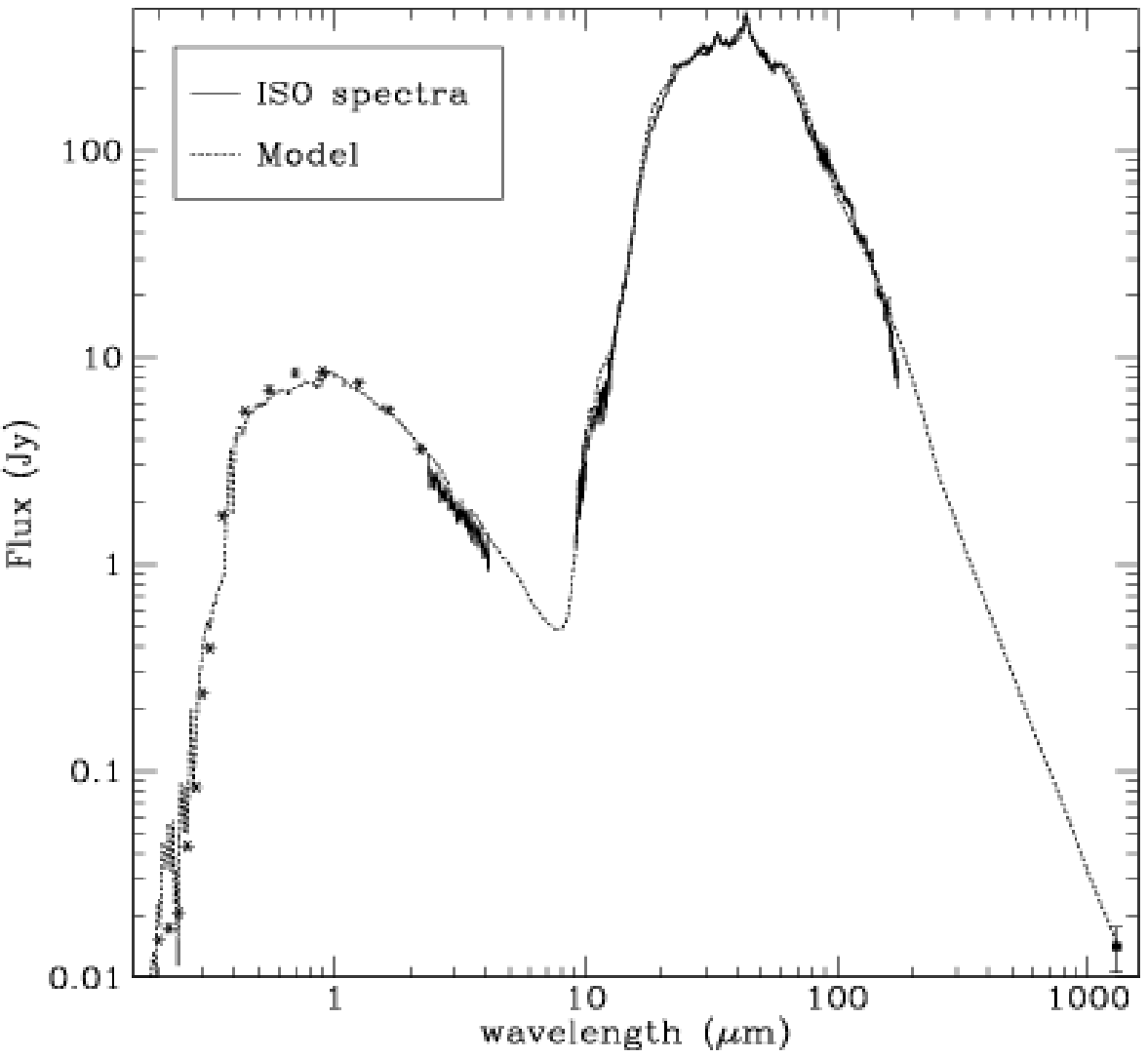} }
  \caption{Spectral energy distribution of 
 {\it classical} post-AGB stars: the C-rich IRAS 19500$-$1709 
(Clube \& Gledhill 2004; left); and the O-rich HD 161796 
(from Hoogzaad et al. 2002; right).}\label{SED1}
\end{figure}

\subsection{Properties of rapidly evolving, heavily obscured post-AGB stars}

 A second class of post-AGB stars is formed by a population of strongly
reddened sources characterized by the conspicuous presence
of a hot dust component in their SEDs attributed to recent or on-going mass
loss (see Figure 3). They are 
distinguished by a narrower galactic distribution, indicative of higher
mass progenitors. Their central stars are usually of B-type suggesting a
fast post-AGB evolution. 
They are surrounded by molecular shells which are in many cases easily 
detectable in CO or OH at (sub-)milimeter and radio wavelengths. In many cases,
shocked-excited molecular hydrogen emission is also detected in 
association with high velocity outflows and/or a strong bipolar morphology. 
A few sources belonging to this group show low excitation nebular 
emission lines, suggesting that the photoionization of the circumstellar 
envelope has already started.

 In extreme cases, these post-AGB stars do not show
any optical counterpart at all; sometimes they are neither detectable at
near-infrared wavelengths (see e.g. Jim\'enez Esteban et al. 2006).
They must represent a population of high-mass stars so 
rapidly evolving to the PN stage that never
become observable as PNe in the optical. 
Most of these sources are O-rich, as expected for stars developing
hot bottom burning (HBB, hereafter) during the previous AGB phase 
(Lattanzio 2003), and show strong OH masers. 
In the infrared they are characterized by the presence of strong 
amorphous silicate absorption features which are in many cases accompanied
by narrow crystalline silicate emission, which is usually interpreted 
as a consequence of a recent phase of  strong mass loss (high-temperature 
annealing; Waters et al. 1996; Sylvester et al. 1999); or, alternatively, as
the result of the presence in the system 
of a long-lived circumbinary disk (low-temperature crystallization; 
Molster et al. 1999). A subgroup of them, the so-called {\it OHPNe}, show 
also radio continuum emission (Garc\'\i a-Hern\'andez et al. these 
proceedings) and may constitute a new class of \textit{infrared PNe}.

\begin{figure} [h!]
\centering
 \resizebox{6cm}{!}{\includegraphics[height=5.8cm,width=5cm]{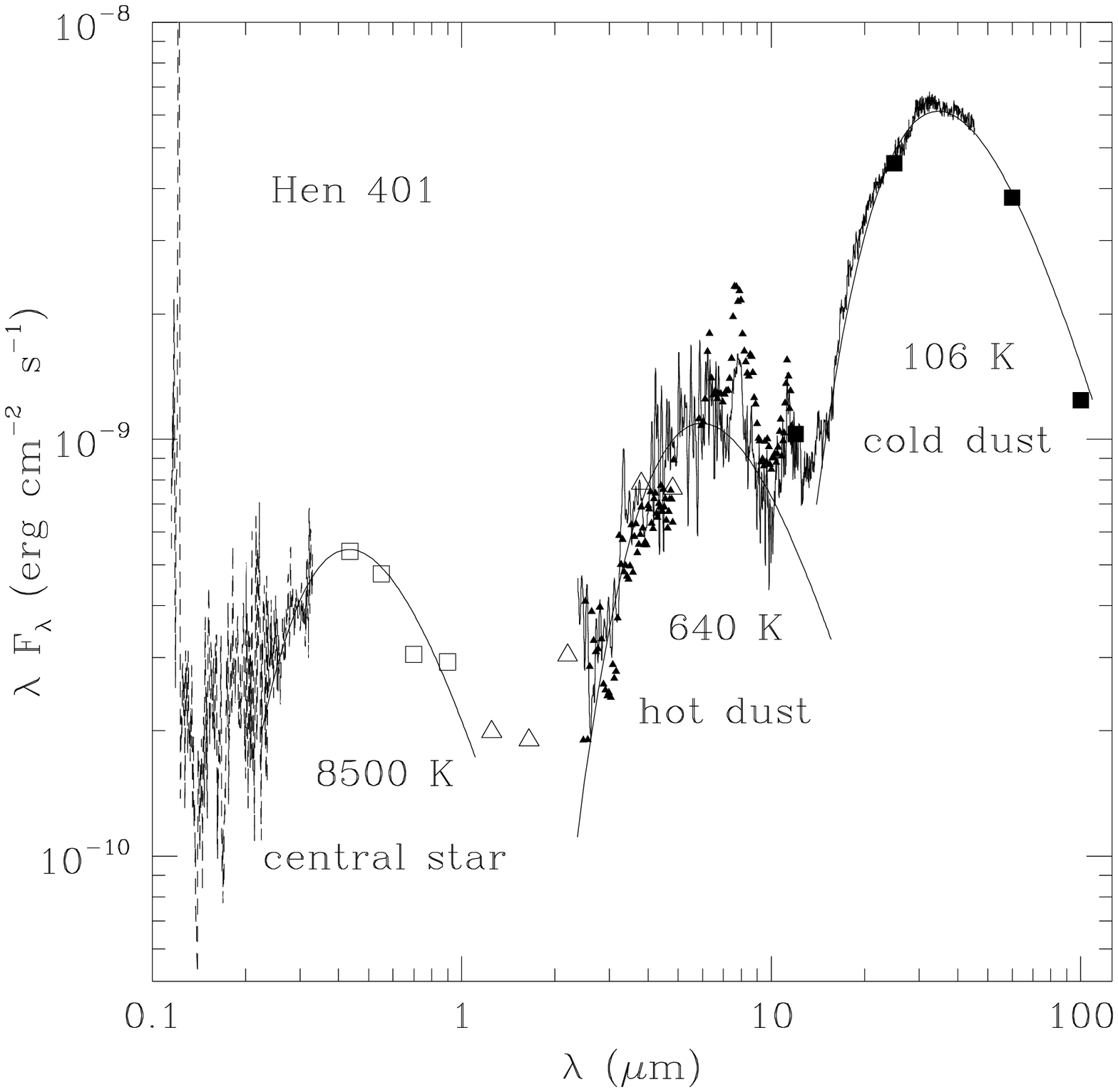} }
 \resizebox{6cm}{!}{\includegraphics[height=5.5cm,width=5cm]{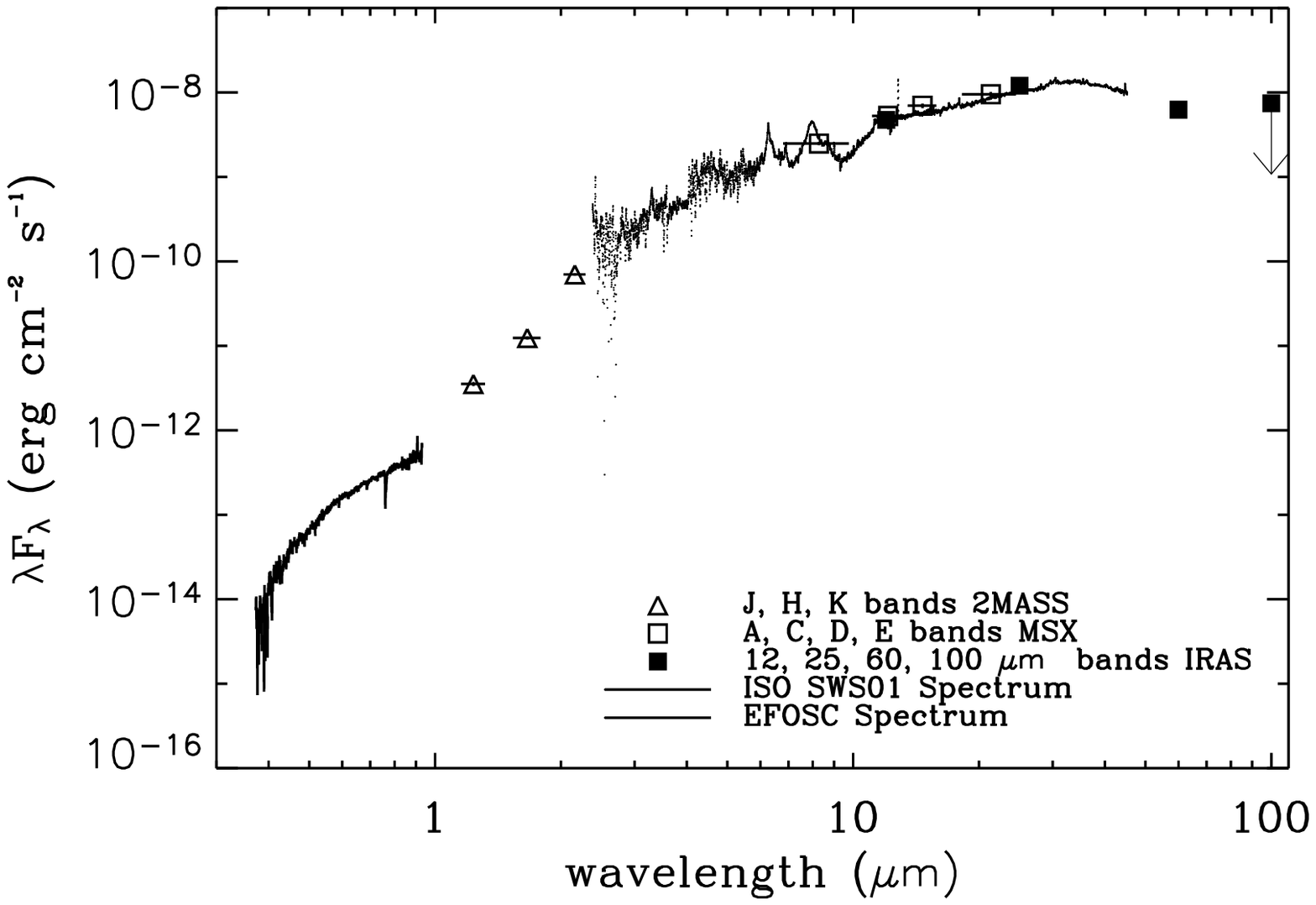} }
  \caption{Spectral energy distribution of rapidly evolving, high mass 
 post-AGB stars. The strongly  bipolar source Hen 3$-$401 
 (from Parthasarathy et al. 2001; left);
 and the heavily obscured, OHPN IRAS 17347$-$3139 (from
 Perea Calder\'on et al. in prep.; right).}\label{SED2}
\end{figure}

\section{The ISO legacy}

Our understanding of the post-AGB transition phase has enormously 
benefited from the results provided by ISO. Based on the analysis of
$\sim$350 ISO spectra of sources spanning the whole evolution from the
AGB to the PN stage (65 of them post-AGB stars), an evolutionary 
scheme has been proposed by Garc\'\i a-Lario \& Perea Calder\'on (2003).
 The scheme takes into account not only the  
shape of the infrared spectrum but also the evolution of the
gas-phase molecular bands and of the solid state features detected in the SWS
spectral range. Two main chemical evolutionary branches are identified
(see Figure 4) which reflect the continuous increase of optical 
thickness in the circumstellar shell of AGB stars, as they evolve towards
the PN stage, as well as the cool down of the envelope as a consequence
of the shell expansion after the end of the strong mass loss phase. Through
these sequences we can also follow in detail  the process of 
condensation and growth of the dust grains formed in the stellar envelope  
until the star becomes a PN. In addition,
there is a clear evolution of carbonaceous material from aliphatic  to
aromatic structures and of the silicates from amorphous to crystalline,
as a consequence of the thermal processing of the grains in the envelope.

\begin{figure}
\centering
 \resizebox{6.5cm}{!}{\includegraphics{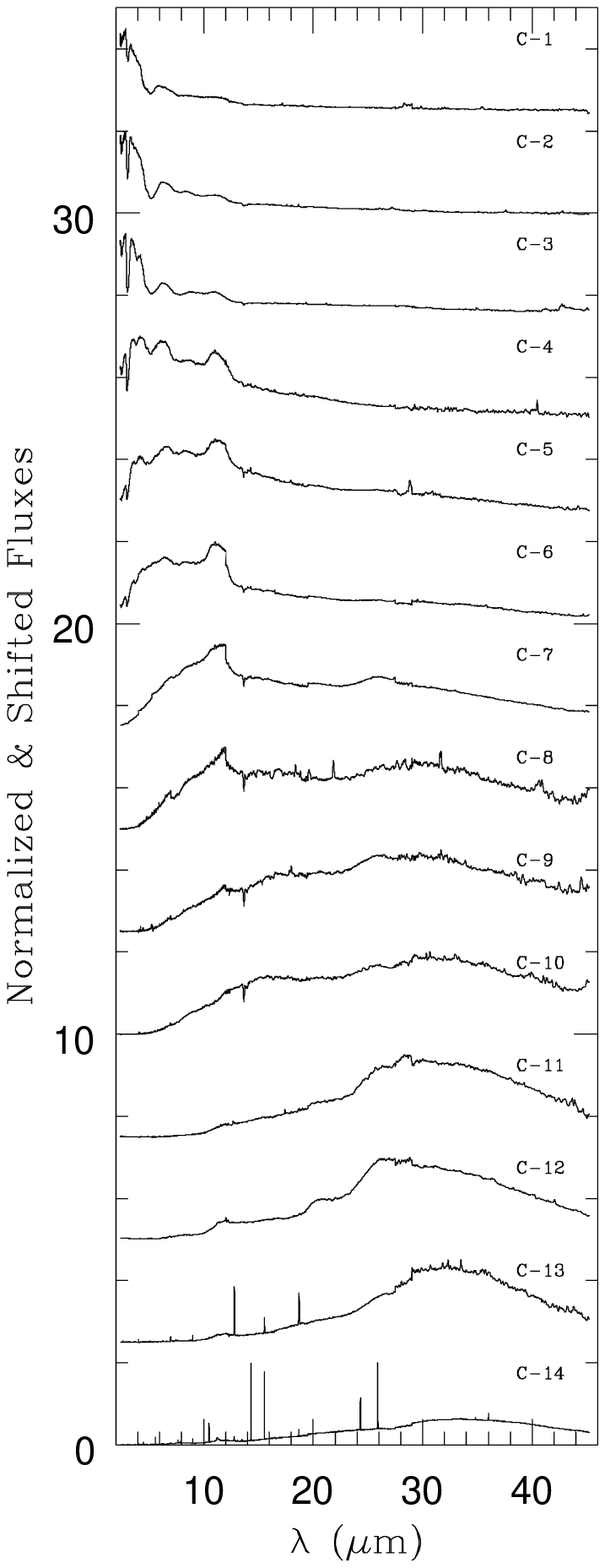} }
 \resizebox{6.5cm}{!}{\includegraphics{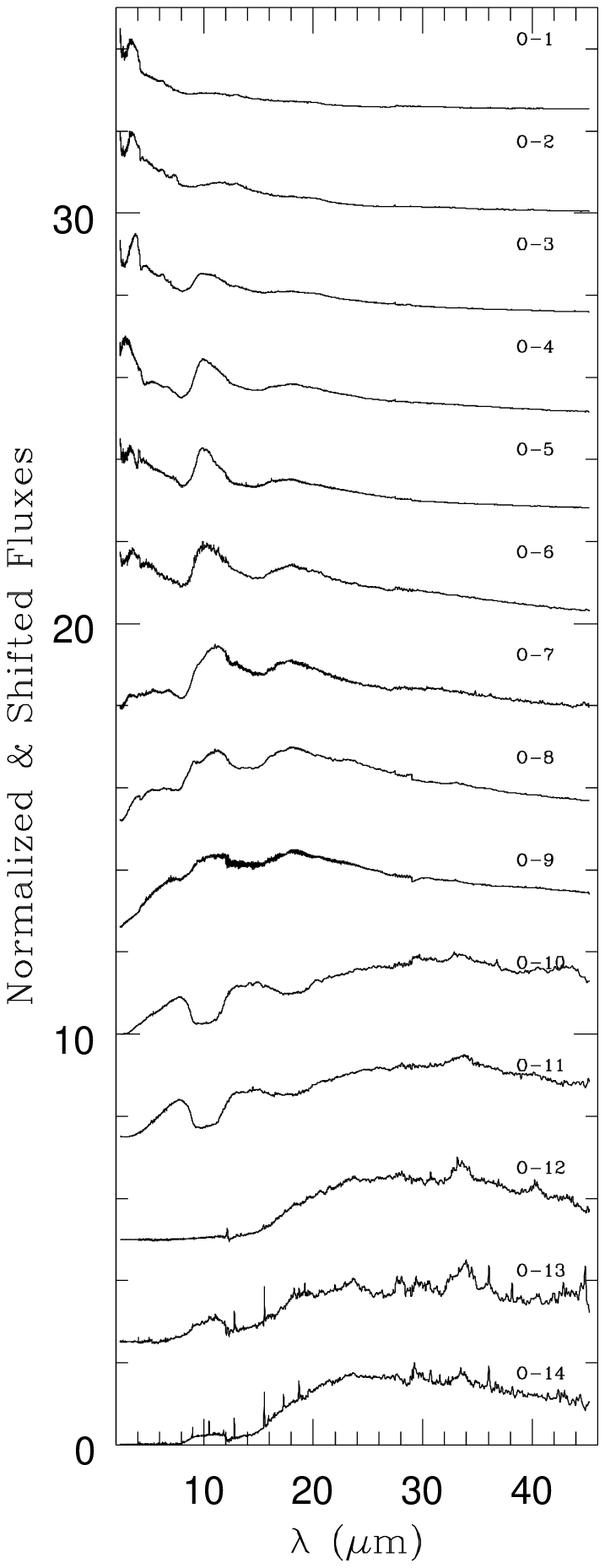} }
  \caption{Spectral sequence followed by intermediate-mass
 C-rich AGB stars (left) and by high-mass 
 O-rich AGB  stars (right) in their way to become PNe (from
 Garc\'\i a-Lario \& Perea Calder\'on 2003). A third sequence (not
 shown) would correspond to the low-mass O-rich stars which do not
 experience the 3rd dredge-up in the AGB, some of which may never 
 become PNe.}\label{isosequence}
\end{figure}

 The ISO spectra of C-rich post-AGB stars show many dust features including 
the well known emission bands at 3.3, 6.2, 7.7, 8.6 and  11.3 $\mu$m generally
attributed to PAHs. These features show different shapes and relative 
intensities as a function of the excitation conditions and grain size 
(Verstraete et al. 2001; Peeters et al. 2002). Additional bands can be 
observed at 11--15 $\mu$m if the PAHs are highly hydrogenated 
(Hony et al. 2001); or at 15--21 $\mu$m, if they are very large in size 
(van Kerckhoven et al. 2000). The 21 $\mu$m feature, only observed so far in 
 C-rich post-AGB stars and in very young PNe has been attributed to 
 various molecular and solid-state species, and the debate is still open. 
 Among the proposed carriers we find
 TiC nanocrystals (von Helden et al. 2000);
 TiC interacting with fullerenes (Kimura et al. 2005); SiC grains 
 (Speck \& Hofmeister 2004); or doped SiC grains (Jiang et al. 2005).
  Also frequently observed in  C-rich 
 post-AGB stars is  the broad and prominent {\it 30 $\mu$m} feature, 
 tentatively identified as MgS (Hony et al. 2002), which is actually a 
 mixture of  two different features centred at 26 and 33 $\mu$m showing a
 variable shape and strength from source to source (Volk et al. 2002).

  In the case of the O-rich AGB stars, the spectral evolution 
 is characterized by the increasing strength of the amorphous silicate 
 emission bands at 9.7 and 18 $\mu$m,  which can eventually turn 
 into absorption when the mass loss rate reaches its maximum value at the end 
 of the thermal pulsing AGB phase. At this moment,  prominent crystalline 
 silicate features take over at 19.8, 23.5, 27.5, 33.5, 40.8, 43.3 and 60 
 $\mu$m which remain visible during the whole post-AGB evolution. They are 
 mostly Mg-rich olivine and pyroxenes (Molster et al. 2002)  but other forms 
 like diopside, enstatite, forsterite (with different relative contents of 
 Mg and Fe) are also detected.  Water ice at 3.1 $\mu$m and at 
 43 and 62 $\mu$m is additionally observed only in the most heavily obscured
 and extremely bipolar sources  (e.g. IRAS 16342$-$3814, Sahai
 et al. 1999;   IRAS 22036+5306, Sahai et al. 2003; or IRAS 17423$-$1755,
 Garc\'\i a-Lario et al. in prep.). The freezeout of water onto dust
 grains requires low temperatures and high densities 
 which are characteristic of the shielded outer regions of the 
 dusty torii seen in the HST images of these stars.

\section{Post-AGB chemistry and stellar evolution}

 Nucleosynthesis models predict the evolution of AGB stars along 
 different chemical paths as a function of the initial mass of the 
 progenitor star. If correct, they should also be able to explain the 
 chemical segregation observed in  post-AGB stars and in PNe.

  Although all stars are born O-rich, reflecting the chemical composition of 
 the ISM, nuclear burning products brought to the stellar 
 surface during the thermal pulsing AGB phase (in the so-called 
 \textit{3rd dredge-up}) can eventually turn the star into C-rich 
 (and s-process enriched) before it enters the post-AGB phase. 
 This 3rd dredge-up is actually expected to achieve its maximum efficiency 
 in stars with masses  in the 1.2 to 3.0 M$_{\odot}$ range, which will 
 become C-rich after a few thermal pulses ending their AGB lifetime as 
 C-rich post-AGB stars. These stars will follow the evolutionary sequence 
 shown in the left panel of Figure 4.  In  contrast, low-mass stars (M$<$1.2 
 M$_{\odot}$) may not experience any significant dredge-up during the 
 AGB phase (Gallino et al. 2004)  and will stay O-rich during the whole 
 AGB-PN evolution. This population of low-mass 
 stars can be identified with the \textit{classical} O-rich
 post-AGB stars not s-process enriched already described in this paper. 
 Above 3-4 M$_{\odot}$, the activation of the HBB 
 will also prevent the transformation of O-rich AGB stars into C-rich 
 sources. The infrared evolution of these high-mass stars  is probably well
 represented by the spectral sequence shown in the right panel of Figure 4.
 Recent observations of heavily obscured O-rich AGB stars confirm that they 
 are actually HBB stars (Garc\'\i a-Hern\'andez et al. these proceedings) and 
 they are suspected to be the progenitors of N-rich type I PNe. 
 Low-mass O-rich AGB stars would follow a similar spectral 
 evolution but in this  case the silicate bands
 may never turn into absorption, as these stars are not expected to 
 develop thick circumstellar shells.

  Metallicity effects cannot be ignored, and actually they
 seem to play an important role in the definition of the mass limits which
 determine the chemical dicotomy observed. 
 In particular, the efficiency of dust production is expected to decrease at
 low metallicities, as well as the  mimimum mass needed for the activation of 
 the HBB or  the number of thermal pulses needed to invert the C/O ratio
 (this would  explain the higher proportion of C-rich PNe detected in  
 the Magellanic Clouds). Implications on AGB lifetimes and on initial/final 
 mass ratios should carefully be considered in future studies. 

\section{The future}

Additional post-AGB samples are expected to become available soon
with the advent of new data from Spitzer and Akari. 
If we can extend the analysis to 
sources located in the Galactic Bulge, the Galactic halo,
the Magellanic Clouds and other galaxies of the Local Group we will
be able for the first time to probe the effect of  
different metallicity environments on the observed evolution.
 Some results were already presented
at this conference based on Spitzer data (like e.g. the detection of the 
first extragalactic post-AGB star in the LMC by Bernard-Salas et al.; or 
the analysis of a few heavily obscured OHPNe in the direction of the 
Galactic Bulge by Garc\'\i a-Hern\'andez et al.),
and we expect to see many more new exciting results in the very near future.
 Determining the stellar yields returned by low- and 
intermediate-mass stars to the ISM as a function of the metallicity 
will certainly improve our knowledge on the chemical evolution of galaxies. 
 High spatial resolution observations in the mid-infrared with narrow 
filters will help to determine the relative distribution of O-rich and
C-rich dust in transition sources showing a dual chemistry and study the  
connection with [WC]-type PNe, where the same phenomenon is observed
(Perea-Calder\'on et al., these proceedings).
Finally, thanks to the improved sensitivity and spatial resolution to be 
provided soon by the Herschel Space Observatory we will be able to 
search for extended far-infrared emission around post-AGB stars. 
Determining the radial density 
distribution of the dust will be essential to recover the mass loss history 
experienced by low- and intermediate-mass stars in the previous AGB phase,
a fundamental ingredient in all evolutionary models.


\begin{thebibliography}{}
\bibitem[Bloecker 1995]{bloecker95}
     {Bl\"ocker, T.} 1995,
     \textit{A\&A} 299, 755
\bibitem[Clube & Gledhill 2003]{clube03}
      {Clube, K.L \& Gledhill, T.M.} 2003,
      \textit{MNRAS} 335, 17
\bibitem[Gallino et al. 2004]{gallino04}
       {Gallino, R., Arnone, E., Pignatari, M. \& Straniero} 2004,
       \textit{MmSAI} 75, 700
\bibitem[Garcia-Hernandez et al. 2002]{garciahernandez02}
       {Garc\'\i a-Hern\'andez, D.A., Manchado, A., Garc\'\i a-Lario et al.}
       2002, \textit{A\&A} 387, 955
\bibitem[Garcia-Lario 1992]{garcialario92}
     {Garc\'\i a-Lario, P.} 1992
     \textit{Ph.D. Thesis}, U. La Laguna, Spain
\bibitem[Garcia-Lario & Perea Calderon 2003]{garcialario03}
        {Garc\'\i a-Lario, P. \& Perea Calder\'on, J.V.} 2003,
        in \textit{Exploiting the ISO Data Archive. Infrared Astronomy in the
        Internet Age}, ESA SP-511, p.97
\bibitem[Garcia-Segura et al. 1999]{garciasegura99}
        {Garc\'\i a-Segura, G., Langer, N., R\'o\.zyczka, M. \& Franco, J.} 
        1999, \textit{ApJ} 517, 767
\bibitem[Hony et al. 2001]{hony01}
        {Hony, S., Van Kerckhoven, C., Peeters, E. et al.} 2001,
	\textit{A\&A} 370, 1030
\bibitem[Hony et al. 2002]{hony02}
        {Hony, S., Waters, L.B.F.M. \& Tielens, A.G.G.M.} 2002,
        \textit{A\&A} 390, 533
\bibitem[Hoogzaad et al. 2002]{hoogzaad02}
         {Hoogzaad, S.N., Molster, F.J., Dominik, C. et al.} 2002,
         \textit{A\&A} 389, 547
\bibitem[Hrivnak et al. 1989]{hrivnak89}
      {Hrivnak, B.J., Kwok, S. \& Volk, K.M.} 1989,
      \textit{ApJ} 346, 265
\bibitem[Hu et al. 1993]{hu93}
      {Hu, J.Y., Slijkhuis, S., de Jong, T. \& Jiang, B.} 1993,
      \textit{A\&AS} 100, 413
\bibitem[Jiang et al. 2005]{jiang05}
      {Jiang, B.W., Zhang, K. \& Li, A.} 2005,
      \textit{ApJ} 630, 77
\bibitem[Jimenez-Esteban et al. 2006]{jimenezesteban06}
        {Jim\'enez-Esteban, F., Garc\'\i a-Lario, P., Engels, D. 
         \& Perea Calder\'on, J.V.} 2006,
        \textit{A\&A} 446, 773
\bibitem[Kelly & Hrivnal 2005]{kelly05}
       {Kelly, D.M. \& Hrivnak, B.J.} 2005,
       \textit{ApJ}  629, 1040       
\bibitem[Kimura et al. 2005]{kimura05}
        {Kimura, Y., Nuth, J.A.III \& Ferguson, F.T.} 2005,
	\textit{ApJ} 632, 159
\bibitem[Kwok et al. 2002]{kwok02}
       {Kwok, S., Volk, K. \& Hrivnak, B.J.} 2002, 
       \textit{ApJ} 573, 720
\bibitem[Lattanzio 2003]{lattanzio03}
        {Lattanzio, J.} 2003,
	in \textit{Planetary Nebulae: Their Evolution and Role in the Universe},
        IAU Symp. 209, eds. S. Kwok, M. Dopita and R. Sutherland, p. 73 
\bibitem[Livio & Soker 1988]{livio88}
        {Livio, M. \& Soker, N.} 1988,
        \textit{ApJ} 329, 764
\bibitem[Molster et al. 1999]{molster99}
	{Molster, F.J., Yamamura, I., Waters, L.B.F.M. et al.} 1999,
        \textit{Nature}, 401, 563
\bibitem[Molster et al. 2002]{molster02}
        {Molster, F.J., Waters, L.B.F.M. \& Tielens, A.G.G.M.} 2002,
	\textit{A\&A} 382, 222
\bibitem[Oudmaijer 1996]{oudmaijer96}
      {Oudmaijer, R.D.} 1996,
      \textit{A\&A} 306, 823
\bibitem[Oudmaijer et al. 1992]{oudmaijer92}
      {Oudmaijer, R.D., van der Veen, W.E.C.J., Waters, L.B.F.M. et al.} 1992,
      \textit{A\&AS} 96, 625
\bibitem[Parthasarathy et al. 2001]{partha01}
       {Parthasarathy, M., Garc\'\i a-Lario, P., Gauba, G. et al.} 2001,
       \textit{A\&A} 376, 941
\bibitem[Peeters et al. 2002]{Peeters02}
         {Peeters, E., Hony, S., Van Kerckhoven, C. et al.} 2002,
	 \textit{A\&A} 390, 1089
\bibitem[Sahai et al. 1999]{sahai99}
        {Sahai R., te Lintel Hekkert, P., Morris, M., Zijlstra, A., \& Likkel, L.} 1999,
        \textit{ApJL} 514, 115
\bibitem[Sahai et al. 2003]{sahai03}
        {Sahai R., Zijlstra, A., S\'anchez Contreras, C. \& Morris, M.} 2003,
         \textit{ApJ} 586, 81
\bibitem[Speck & Hofmeister 2004]{speck04}
        {Speck, A.K. \& Hofmeister, A.M.} 2004,
	\textit{ApJ} 600, 986
\bibitem[Suarez et al. 2006]{suarez06}
       {Su\'arez, O., Garc\'\i a-Lario, P., Manchado, A. et al.} 2006,
       \textit{A\&A} (submitted)
\bibitem[Sylvester et al. 1999]{sylvester99}
       {Sylvester, R., Kemper, F., Barlow, M.J. et al.} 1999,
       \textit{A\&A} 352, 587
\bibitem[Ueta et al. 2000]{ueta00}
	{Ueta, T., Meixner M. \& Bobrowsky, M.} 2000
        \textit{ApJ} 528, 86
\bibitem[Van Kerkhoven et al. 2000]{vankerkhoven00}
        {Van Kerkhoven, C., Hony, S., Peeters, E. et al.} 2000,
	\textit{A\&A} 357, 1013
\bibitem[van Winckel (2003)]{vanWinckel03}
     {van Winckel, H.} 2003, 
     \textit{Ann.Rev.A\&A} 41, 391
\bibitem[Verstraete et al. 2001]{verstraete01}
      {Verstraete, L., Pech, C., Moutou, C. et al.} 2001,
      \textit{A\&A} 372, 981
\bibitem[Volk et al. 2002]{volk02}
     {Volk, K., Kwok, S., Hrivnak, B.H. \& Szczerba, R.} 2002,
     \textit{ApJ} 567, 412
\bibitem[von Helden et al. 2000]{vonhelden00}
      {von Helden, G., Tielens, A.G.G.M., van Heijnsbergen, D. et al.} 2000,
      \textit{Science} 288, 313
\bibitem[Waelkens \& Waters 2004]{waelkens04}
     {Waelkens, C. \& Waters, L.B.F.M.} 2004,
      in \textit{Asymptotic Giant Branch Stars}, H.J. Habing and H. Olofsson 
      (eds.) Springer-Verlag, New York
\bibitem[waters et al. 1996]{waters96}
      {Waters, L.B.F.M., Molster, F.J, de Jong, T. et al.} 1996,
      \textit{A\&A} 315, L361      
\end{thebibliography}
\end{document}